\documentclass[preprint]{aastex}

\usepackage{natbib}
\usepackage{amsmath} \usepackage{amssymb}
\usepackage[colorlinks=true,urlcolor=black,linkcolor=black,citecolor=black]{hyperref}

\shorttitle{Nonparametric Forecasts of the CMB Angular Power Spectra for the Planck Mission}
\shortauthors{Aghamousa, Arjunwadkar, and Souradeep}

 \usepackage{xcolor}
 \newcounter{attnctr} 

\begin{document}

\title{Nonparametric Forecasts of the CMB Angular Power Spectra for the Planck Mission}

\author{Amir Aghamousa}
\affil{Centre for Modeling and Simulation, University of Pune, Pune 411 007 India}
\affil{Department of Physics, University of Pune, Pune 411 007 India}
\email{amir@cms.unipune.ac.in}

\author{Mihir Arjunwadkar}
\affil{National Centre for Radio Astrophysics, University of Pune Campus, Pune 411 007 India}
\affil{Centre for Modeling and Simulation, University of Pune, Pune 411 007 India,}
\email{mihir@ncra.tifr.res.in}

\and

\author{Tarun Souradeep}
\affil{Inter-University Centre for Astronomy and Astrophysics, University of Pune Campus, Pune 411 007 India}
\email{tarun@iucaa.ernet.in}

\begin{abstract}
The Planck mission, designed for making measurements of the cosmic microwave background (CMB) radiation with unprecedented accuracy and angular resolution, is expected to release its first data set in the near future.
For the first time in the CMB history, extensive measurements of the CMB polarization will be made available for the entire sky.
Such precise and rich data are expected to contain a great wealth of information about the Universe.
The information in the CMB data is conveniently represented in terms of angular power spectra for temperature and polarization.
A proper estimation of these CMB power spectra from data is the first step in making inferences about the Universe and,
in particular, cosmological parameters that govern the Universe.
In this paper, we provide forecasts for the $TT$, $EE$, and $TE$ angular power spectra for the Planck mission.
Our forecasts are made using synthetic data based on the best-fit $\Lambda$CDM model
while conforming to the characteristics and parameters of the Planck mission.
We have analysed such synthetic data sets using a nonparametric function estimation formalism that is otherwise asymptotically model-independent.
Indeed, the power spectra estimated from such synthetic data turn out to be sufficiently close to the corresponding $\Lambda$CDM spectra.
Our analysis further indicates that the three angular power spectra are determined sufficiently well for $2 \le l \lesssim l_{max}$,
where $l_{max} = 2500 \mbox{ ($TT$), } 1377 \mbox{ ($EE$), and } 1727 \mbox{ ($TE$)}$ respectively.
A signature of reionization is evident in the $EE$ and $TE$ power spectra in the form of a bump at low $l$s.
Nonparametric confidence bands
in the phase shift ($\phi_m$) vs.\ acoustic scale ($l_A$) plane,
corresponding to the first eight peaks in the $TT$ power spectrum,
show a confluence region for $300\lesssim l_A \lesssim 305$,
which is in good agreement with the estimate $l_A=300$ based on the best-fit $\Lambda$CDM model.
From our results, we expect that the real Planck data, when released, should lead to accurate model-independent estimates of CMB angular power spectra using our nonparametric regression formalism.
\end{abstract}


\section{Introduction}
\label{introduction}
Observation of the microwave sky reveals that the temperature of the cosmic microwave background (CMB) radiation is not exactly the same in all directions.
These small fluctuations in the temperature are imprinted on the entire sky, implying that the CMB is anisotropic.
These primordial anisotropies were first discovered in 1992 by the COsmic Background Explorer (COBE).
This was followed up by a remarkable series of ground-based and balloon-borne experiments,
and most recently by the Wilkinson Microwave Anisotropy Probe (WMAP).
These fluctuations are believed to have been generated within $10^{-35}$ seconds of the Big Bang.
CMB anisotropies are therefore a rich source of information about the early Universe, and have revolutionized the way we understand our Universe.
A study of CMB anisotropies also helps in probing the fundamental physics at energy scales much higher in magnitude compared to those accessible to particle accelerators.
CMB anisotropies are sensitive to the classical cosmology parameters such as expansion rate, curvature, cosmological constant, matter content, radiation content, and baryon fraction, and provide insights for modeling structure formation in the Universe \citep{1996LNP...470..207H}.
For example, measurements of the CMB anisotropies with ever-increasing precision have made it possible to establish a standard cosmological model that asserts that the Universe is nearly spatially flat \citep{Smooth_Nobel}.

The CMB contains a far greater wealth of information about the Universe through its (linear) polarization.
The CMB has acquired linear polarization through Thomson scattering during either decoupling or reionization,
sourced by the quadrupole anisotropy in the radiation distribution at that time \citep{Rees1968, Hu_White_1997}.
This quadrupole anisotropy could originate in different sources \citep{Hu_White_1997}:
the scalar or density fluctuations, vector or vortical fluctuations, and tensor or gravitational wave perturbations.
Scalar mode represents the fluctuations in energy density of the cosmological plasma at the last scattering
that causes a velocity distribution which leads to blue-shifted photons.
Vector mode represents the vorticity on the plasma which causes Doppler shifts that result into quadrupolar lobes.
However, such vorticity will be damped by expansion (as are all motions that are not enhanced by gravity),
and its effect are expected to be negligible.
Tensor perturbation are the effect of gravity waves which stretch and squeeze space in orthogonal directions.
This also stretches the photon wavelength, and hence produces a quadrupolar variation in temperature.

The dependence of CMB polarization on cosmological parameters differs from that of temperature anisotropies.
As such, it provides additional constraints on cosmological parameters that help break degeneracies.
Accurate measurements of CMB polarization will therefore enable us to affirm the validity/consistency of different cosmological models \citep{Zaldarriaga2004, Zaldarriaga_Spergel_Seljak_1997, Eisenstein1999}.
CMB polarization information is further expected to help determine initial conditions and evolution of structure in the Universe, origin of primordial fluctuations, existence of any topological defects, composition of the Universe, etc.

The Planck mission \citep{TPC2006} is a space-based full-sky probe for third-generation CMB experiments designed for this purpose.
Indeed, one of the main objectives of this mission is to measure the primordial fluctuations of the CMB with an accuracy prescribed by the fundamental astrophysical limits, through improvements in sensitivity and angular resolution, and through better control over noise and confounding foregrounds.
The higher angular resolution of Planck implies that higher-order peaks in the CMB angular spectra can be determined with better precision, which in turn translates to determination of cosmological parameters (such as baryon and dark matter densities) with improvement in statistical precision by an order of magnitude.
Furthermore, the Planck mission is designed to make extensive measurements of the $E$-mode polarization spectrum over multipoles up to $l \approx 1500$ with unprecedented precision, together with good control over polarized foreground noise.
These measurements are expected to provide insights into the physics of early Universe, epoch of recombination, structure formation,
allowable modes of primordial fluctuations (adiabatic vs.\ isocurvature modes), reionization history of the Universe, and help in establishing constraints on the primordial power spectrum.
Planck will also help constrain the fundamental physics at high energies that are not possible to probe through terrestrial experiments \citep{TPC2006}.

In our previous work \citep{AAS2012}, we estimated the CMB $TT$ power spectrum from four WMAP data realizations using a nonparametric function estimation methodology \citep{Beran2000,GMN+2004}.
This methodology does not impose any specific form or model for the power spectrum,
and determines the fit by optimizing a measure of smoothness that depends only on characteristics of the data.
This ensures that the fit and the subsequent analysis is approximately model-independent for sufficiently large data sizes. 
Further, this methodology quantifies the uncertainties in the fit in the form of a high-dimensional ellipsoidal confidence set that is centered at this fit and captures the true but unknown power spectrum with a pre-specified probability.
This confidence set is the prime inferential object of this methodology which allows addressing complex inferential questions about the data meaningfully in a unified framework.

The Planck mission is expected to release high-accuracy CMB data sets in the near future.
In this paper, we therefore attempt to forecast the three CMB power spectra for the CMB to see what can be expected (and inferred) from real Planck data when it gets released, when analysed using this nonparametric regression methodology.
For this purpose,
we use synthetic Planck-like data conforming to the specifications and parameters of the Planck mission.
This synthetic data is based on the assumption that the best-fit $\Lambda$CDM model is the true model of the Universe.

In what follows, Sec.\ \ref{T_P_PS} briefly describes the three CMB angular power spectra, followed by a description  (Sec.\ \ref{simulating}) of the synthetic data used in this work.
This is followed by the results (Sec.\ \ref{results}) and a conclusion (Sec.\ \ref{conclusion}).

\section{Temperature and polarization power spectra}
\label{T_P_PS}
Polarized radiation is typically characterized in terms of the Stokes parameters $I, Q, U$ and $V$ \citep{Jackson1998}.
The parameter $V$ describing circular polarization is not generated by Thomson scattering for the CMB.
The two Stokes parameters $Q$ and $U$ describing linear polarization form the components of a rank-2 symmetric trace-free tensor $\mathcal{P}_{ab}$.
Since any two-dimensional symmetric tensor can be represented using two scalar fields,
$\mathcal{P}_{ab}$ can be represented using to the \emph{electric} (i.e., gradient) $P_E$ and the \emph{magnetic} (i.e., curl) $P_B$ modes of polarization.
This decomposition is unique, and is similar to the decomposition of a vector field into a gradient and a divergence-free vector field \citep{Challinor2004}.
The $P_E$ and $P_B$ scalars are defined on a sphere, and can therefore be expanded in a spherical harmonics basis as
\begin{eqnarray}
P_E(\theta,\phi) & = & \sum_{l\geq2} \sum_{|m|\leq l} \sqrt{\frac{(l-2)!}{(l+2)!}} a_{lm}^E Y_{lm}(\theta,\phi) \\
P_B(\theta,\phi) & = & \sum_{l\geq2} \sum_{|m|\leq l} \sqrt{\frac{(l-2)!}{(l+2)!}} a_{lm}^B Y_{lm}(\theta,\phi),
\end{eqnarray}
which define the $E$- and $B$-mode multipoles $a_{lm}^E$ and $a_{lm}^B$ respectively \citep{Challinor2004, TPC2006, Hu_White_1997}.
Any CMB measurement can be decomposed into three maps ($T$, $E$ and $B$ respectively).
A total of six angular power spectra ($TT$, $EE$, $BB$, $TE$, $TB$, and $EB$) can be obtained from these three components \citep{Oliveira2005}.
These six power spectra are defined by expanding the $T$, $E$ and $B$ maps in terms of spherical harmonics,
resulting into the following correlation structure:
\begin{equation}
<a_{lm}^{Y*} a_{l^\prime m^\prime}^{Y^\prime}> = C_l^{YY^\prime} \delta_{ll^\prime} \delta_{mm^\prime},
\end{equation}
where $Y$, $Y^\prime$ are $E$, $B$ or $T$.
In the absence of parity violation, and under the assumption of Gaussian fluctuations,
the temperature and polarization anisotropies of the CMB are described by the
$C_l^{TT}, C_l^{EE},C_l^{BB},C_l^{TE}$ power spectra completely \citep{Challinor2004, Hu_White_1997, Zaldarriaga2004}.
Since the $B$-mode polarization is not expected to be detected by the Planck mission accurately \citep{TPC2006},
we focus only on the $C_l^{TT}$, $C_l^{EE}$ and $C_l^{TE}$ power spectra.

The standard deviation of the $C_l^{TT}, C_l^{EE},C_l^{TE}$ power spectra is approximately given by \citep{TPC2006, knox1995}
\begin{eqnarray}
\label{var_cltt}
(\Delta C_l^{TT}) &\simeq& \frac{2}{(2l+1)f_{sky}} (C_l^TT + \omega_T^{-1}W_l^{-2})^2 \\
\label{var_clee}
(\Delta C_l^{EE}) &\simeq& \frac{2}{(2l+1)f_{sky}} (C_l^EE + \omega_P^{-1}W_l^{-2})^2 \\
\label{var_clte}
(\Delta C_l^{TE}) &\simeq& \frac{2}{(2l+1)f_{sky}} \left((C_l^EE + \omega_P^{-1}W_l^{-2}) (C_l^TT + \omega_T^{-1}W_l^{-2}) + (C_l^TE)^2 \right),
\end{eqnarray}
where $\omega_T= (\sigma_{p,T} \theta_{FWHM})^{-2}$, and $\omega_P= (\sigma_{p,P} \theta_{FWHM})^{-2}$ are the weights per solid angle for temperature and polarization, and $f_{sky}$ is the fraction of observed sky in the experiment.
The $\sigma_{p,T}$ and $\sigma_{p,P}$ are noise standard deviations per resolution element ($\theta_{FWHM} \times \theta_{FWHM}$).
The window function for a Gaussian beam is
\begin{equation}
 W_l =\exp\left(-l(l+1)/(2l_{beam}^2)\right),
 \label{gaussian_beam}
\end{equation}
where $l_{beam}=\sqrt{8\ln 2}(\theta_{FWHM})^{-1}$.

The importance of decomposing the $\mathcal{P}_{ab}$ tensor in terms of the $E$ and $B$ modes comes from the fact that linear scalar perturbations do not generate any $B$-mode polarization \citep{Hu_White_1997}.
The tensor mode contributes to both $E$ as well as $B$ modes, whereas the vector mode contributes only to the $B$-mode polarization.
Therefore the $E$ part of the decomposition stems from the scalar and tensor modes, and the $B$ part originates in the vector and tensor modes \citep{TPC2006, Challinor2004, Hu_White_1997}.


Cosmological relevance of these polarization modes is as follows.
The $E$-mode mainly follows the velocity of the cosmological plasma at decoupling.
Compared to the temperature anisotropies, which originate in photon density fluctuations at the last scattering,
the $E$ mode therefore contains more information about some of the cosmological parameters
\citep{Zaldarriaga_Spergel_Seljak_1997, Challinor2004}, and can lead to better estimates of parameters
such as the  baryon and cold dark matter densities.
Polarization power spectra have an oscillatory structure that is analogous to that of the $TT$ power spectrum.
For example, peaks in $EE$ power spectrum are out of phase with those in the $TT$ power spectrum due to anisotropy generated at the last scattering.
The $TE$ power spectrum, which has a higher amplitude compared to the $EE$ power spectrum, is a measure of the correlations (positive or negative) between density and velocity fluctuations \citep{Scott_Smoot2006, Challinor2004}.
The phase difference between acoustic peaks in the $TT$, $EE$ and $TE$ power spectra can be used as a model-independent check for the physics of acoustic oscillations \citep{Challinor2004}.

Adiabatic and isocurvature perturbations also have different effects on the phase of the polarization spectra:
Predicted polarization power spectra for isocurvature perturbations show out-of-phase peaks and dips
compared to those for adiabatic perturbations such that the power spectra from isocurvature perturbations appear to be $l$-shifted versions of those for adiabatic perturbations \citep{Sievers2007}.
A meaningful estimation of the polarization power spectra can therefore be used to determine which of the two scenarios is closer to truth.

Another cosmological phenomenon that affects the polarization spectra is reionization:
Reionization of Universe started when the first generation of stars started producing a flux photons.
The resulting free electrons started re-scattering the CMB radiation.
Although only a small fraction of CMB photons got scattered this way during the reionization era,
the imprints of reionization are expected to be seen as distortions in the polarization power spectra at large angular scales of the order of 10 degrees.
The height and location of the reionization bump \citep{Zaldarriaga1997, Kaplinghat2003, Holder2003} expected at low multipoles ($l\lesssim 20$) has information
related to total optical depth and the reionization epoch redshift \citep{Zaldarriaga_Spergel_Seljak_1997, Kaplinghat2003}.
Although the precision of reionization bump detection is limited by cosmic variance at low $l$ \citep{Hu_Holder_2003},
constraining it will help understand the reionization history better,
and break degeneracies between several cosmological parameters by constraining the optical depth better \citep{Zaldarriaga_Spergel_Seljak_1997, Eisenstein1999}.

 \begin{figure}
  \centerline{\includegraphics[height=0.4\textheight]{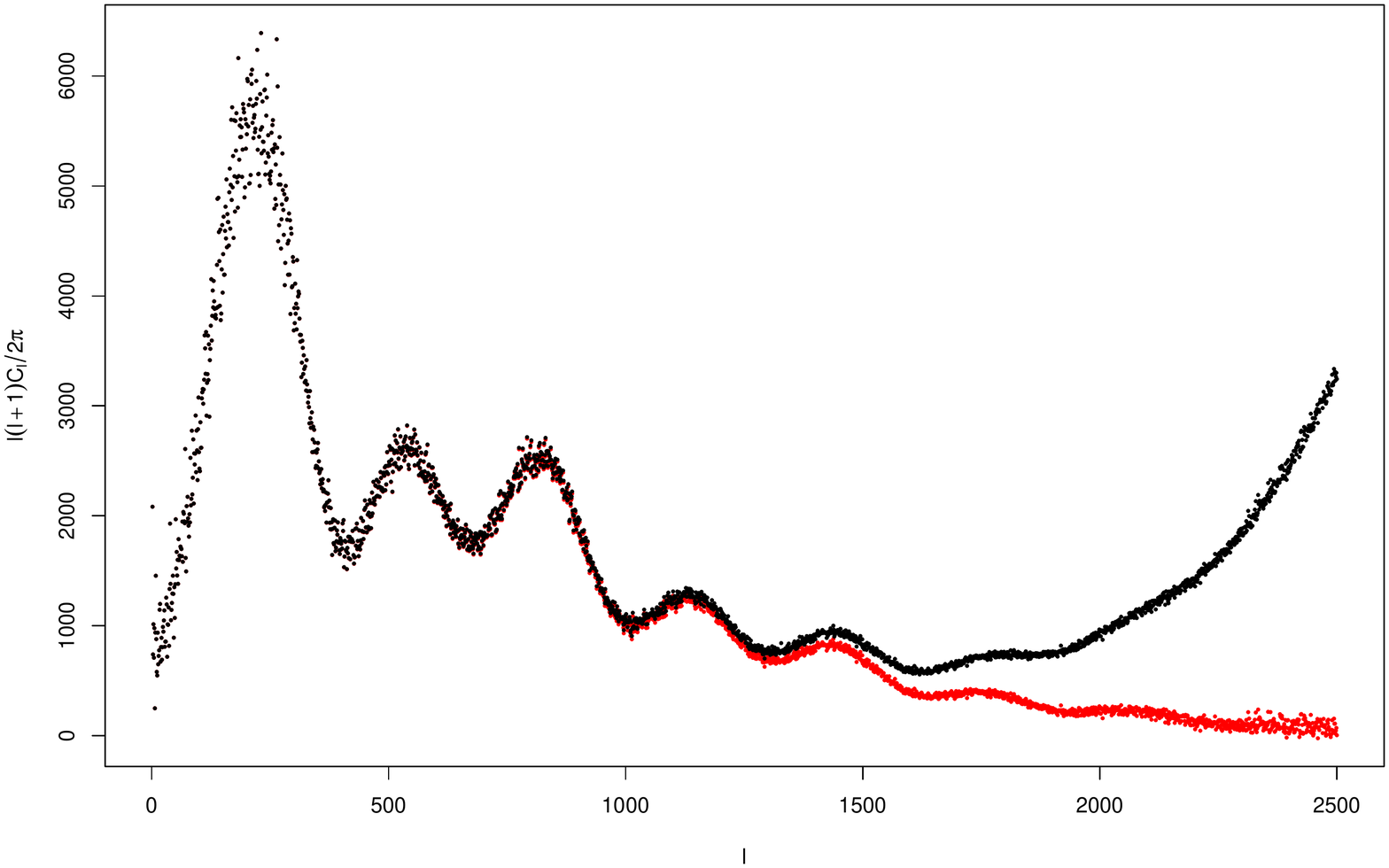}}
  \caption{\label{fig_TT_sim}A realization of simulated $TT$ power spectrum data for the Planck mission, generated using FuturCMB \citep{futurcmb}. Black points: data including noise; red points: simulated data after subtracting noise.}
 \end{figure}
 \begin{figure}
  \centerline{\includegraphics[height=0.4\textheight]{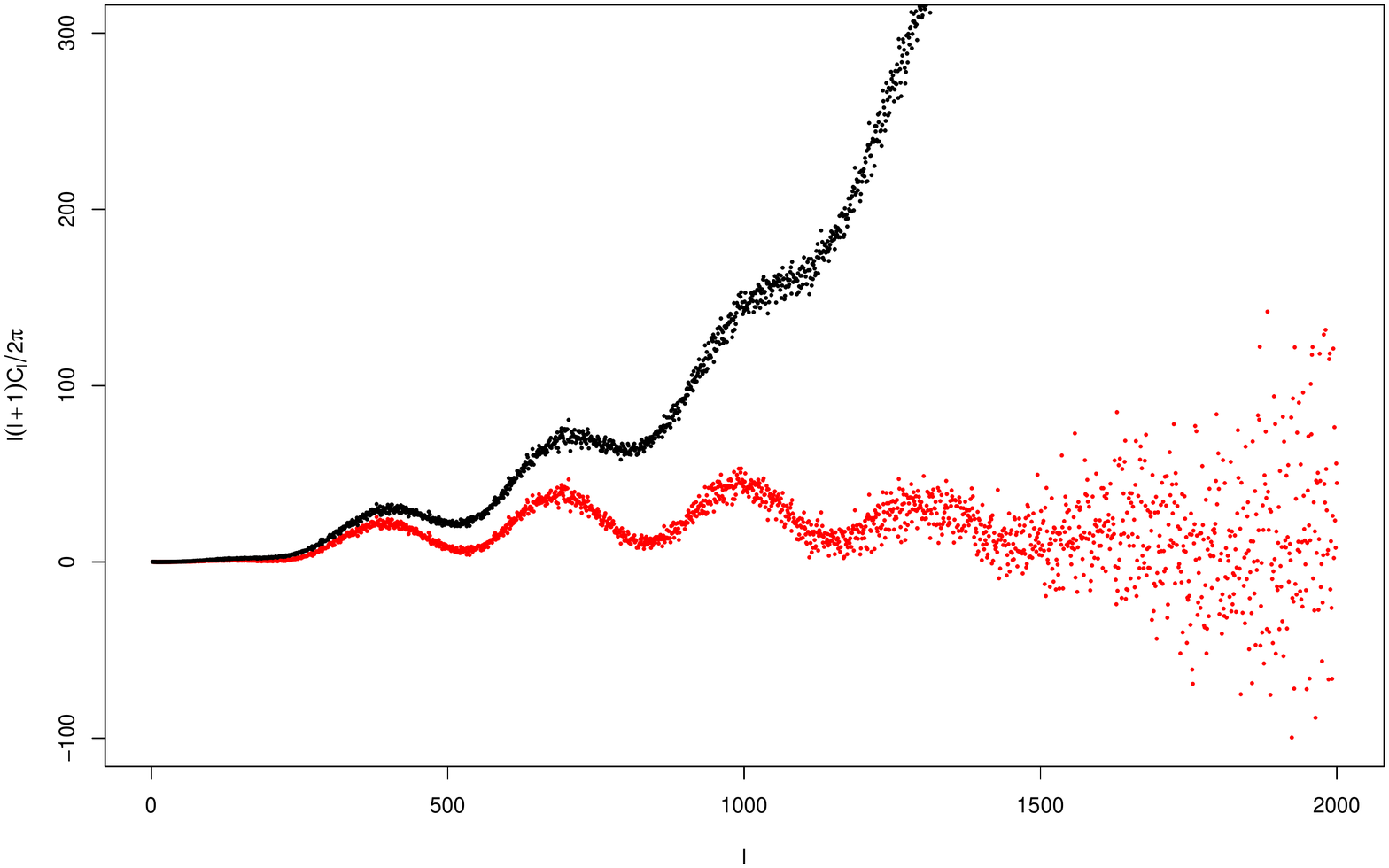}}
  \caption{\label{fig_EE_sim}A realization of simulated $EE$ power spectrum data for the Planck mission, generated using FuturCMB \citep{futurcmb}. Black points: data including noise; red points: simulated data after subtracting noise.}
 \end{figure}

 \begin{figure}[h]
  \centerline{\includegraphics[height=0.4\textheight]{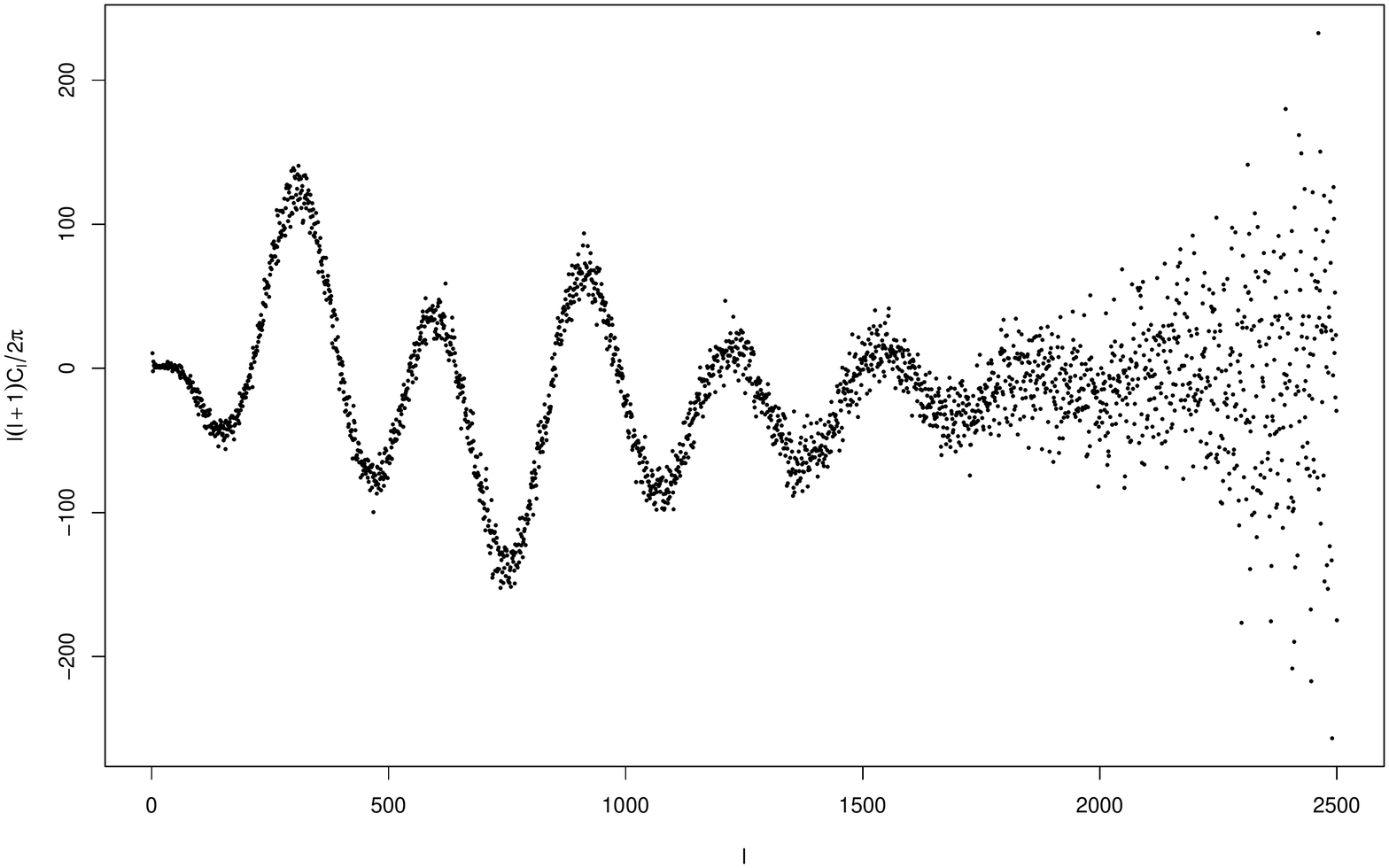}}
  \caption{\label{fig_TE_sim}A realization of simulated $TE$ power spectrum data for the Planck mission, generated using FuturCMB \citep{futurcmb}. FuturCMB assumes the noise to be zero for the $TE$ data.}
 \end{figure}

\section{Synthetic data for the Planck mission}
\label{simulating}

For forecasting the CMB angular power spectra for the Planck mission,
we generate the synthetic Planck-like data using the FuturCMB code \citep{futurcmb}.
FuturCMB generates a simulated angular power spectrum using a user-provided theoretical power spectrum $C_l^{true}$ (which is assumed to be the true spectrum)
for frequency channels representing Planck measurements, and generates the corresponding noise power spectrum $N_l$ conforming to the Planck characteristics.
This is done by generating a random realization of the spherical harmonics coefficients $a_{lm}$, assumed to be Gaussian random variables with
mean zero and variance
\begin{equation}
\label{var_alm}
\text{Var}(a_{lm}) = C_l^{true} + N_l,
\end{equation}
where $C_l^{true}$ is a proxy for the true but otherwise unknown angular power spectrum.
For $C_l^{true}$, we use spectra generated using CAMB \citep{Lewis:1999bs} for the best-fit $\Lambda$CDM cosmological parameters
obtained from the WMAP 7-year data.
We also limit FuturCMB to $l\leq 2500$, a range that corresponds to the three Planck frequency channels (100, 142 and 217 GHz).
$N_l$ is the noise power spectrum given by
\begin{equation}
N_l = \omega^{-1} W_l^{-2},
\end{equation}
where $\omega$ is $\omega_T$ and $\omega_P$ for temperature and polarization respectively,
and $W_l$ is the window function for a Gaussian beam (Eq.\ \ref{gaussian_beam}).
The noise in the $TE$ power spectrum is taken to be zero because noise contributions from different maps are uncorrelated \citep{futurcmb}.
FuturCMB then calculates the power spectra data $C_l^{map}$ as
\begin{equation}
\label{cl_map}
C_l^{map} = \frac{1}{2(l+1)} \sum_{m=-l}^{+l} |a_{lm}|^2.
\end{equation}
Eq.\ \ref{cl_map} is an unbiased estimator of Eq.\ \ref{var_alm}; its expected value is therefore equal to $C_l^{true} + N_l$.
The noise spectra are expected to dominate over the true spectrum for sufficiently high values of $l$.
This is seen in figures \ref{fig_TT_sim}--\ref{fig_TE_sim},
where the black points represent the FuturCMB output $C_l^{map}$ of the $TT$, $EE$ and $TE$ spectra:
the upward trends in the tail of $TT$ and $EE$ spectra are the result of noise dominating the data at high $l$s.
The $TE$ power spectrum, on the other hand, does not show any such upward trend because the noise in the $TE$ spectrum is assumed zero.
To obtain synthetic $TT$ and $EE$ data, we therefore subtract the corresponding noise spectra from the $C_l^{map}$ output of FutureCMB (Figures \ref{fig_TT_sim}--\ref{fig_EE_sim}, red points).
The covariance matrix of the simulated angular power spectra is taken to be a diagonal matrix with diagonal elements defined in Eq.\ \ref{var_cltt}, \ref{var_clee} and \ref{var_clte} for $TT$, $EE$, and $TE$ respectively.

\section{Results and discussion}
\label{results}

 \begin{figure}
  \centerline{\includegraphics[height=0.38\textheight]{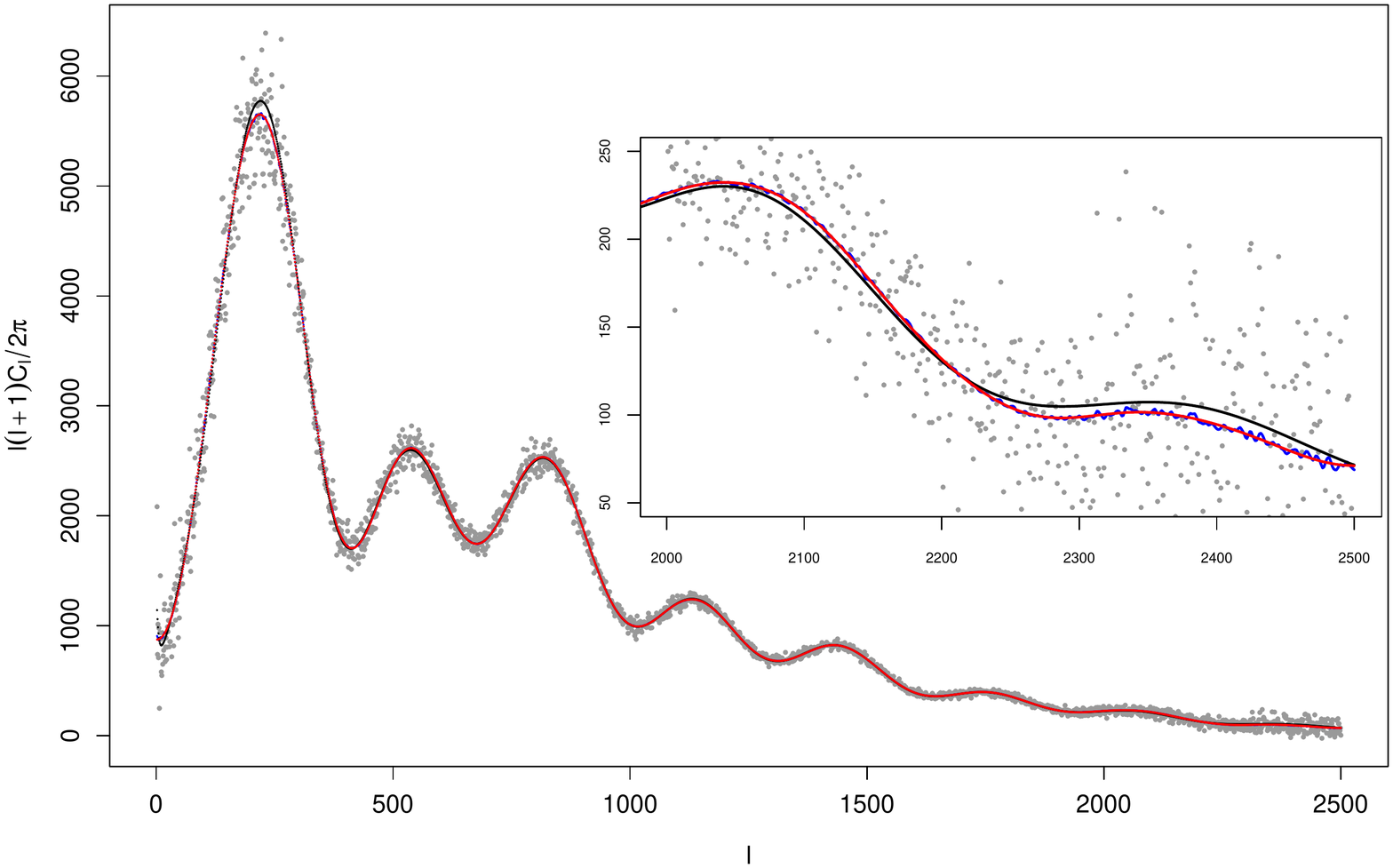}}
  \caption{\label{fig_TT_comp} $TT$ nonparametric fits. Blue, full-freedom fit (EDoF$\approx$72); red, restricted-freedom fit (EDoF$=$27); black: best-fit $\Lambda$CDM spectrum; grey: simulated data realization.}
 \end{figure}
 \begin{figure}
  \centerline{\includegraphics[height=0.38\textheight]{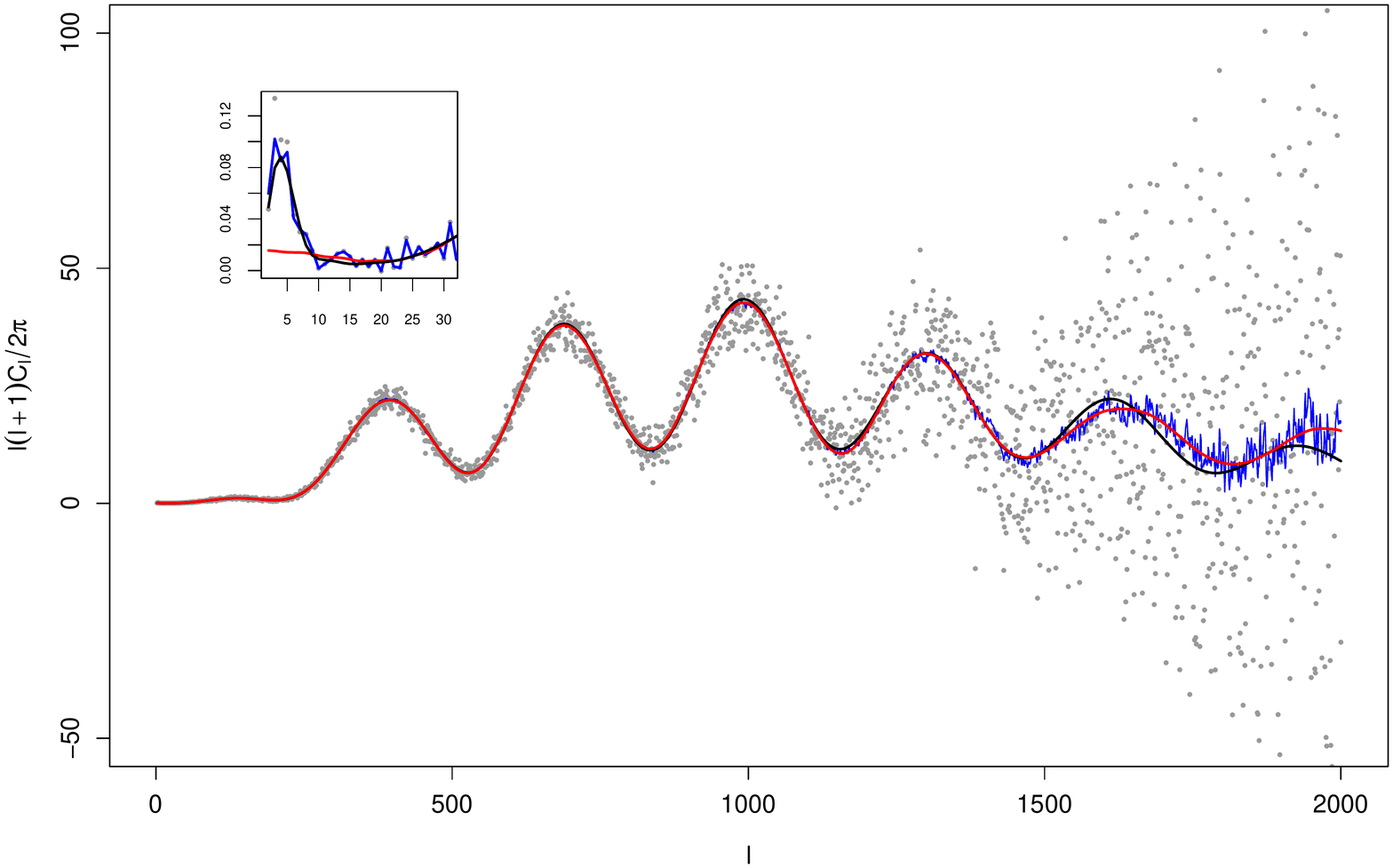}}
  \caption{\label{fig_EE_comp} $EE$ nonparametric fits. Blue, full-freedom fit (EDoF$\approx$190); red, restricted-freedom fit (EDoF$=$24); black: best-fit $\Lambda$CDM spectrum; grey: simulated data realization.}
 \end{figure}

 \begin{figure}[h]
  \centerline{\includegraphics[height=0.38\textheight]{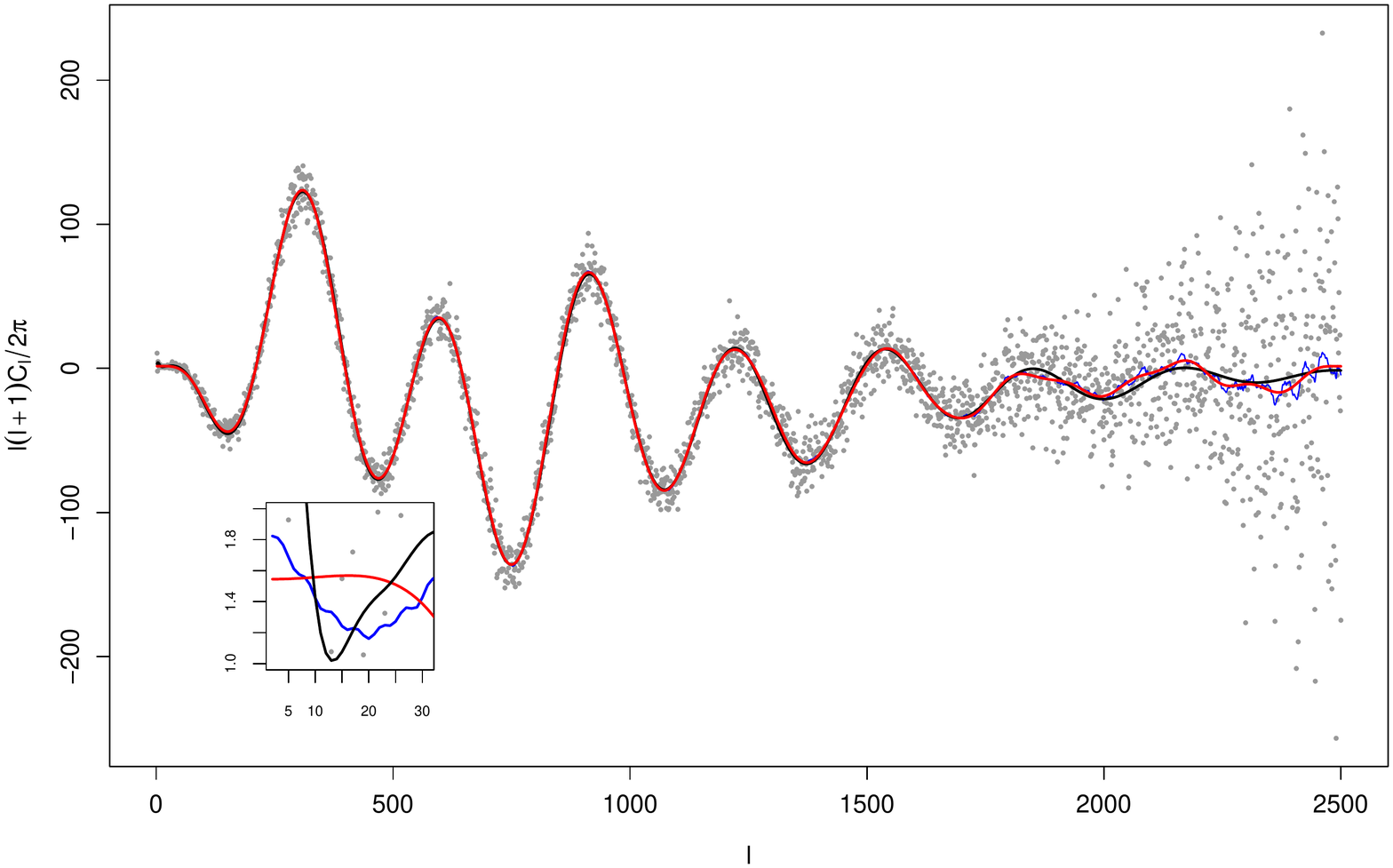}}
  \caption{\label{fig_TE_comp} $TE$ nonparametric fits. Blue, full-freedom fit (EDoF$\approx$95); red, restricted-freedom fit (EDoF$=$40); black: best-fit $\Lambda$CDM spectrum; grey: simulated data realization.}
 \end{figure}

\paragraph{Nonparametric fits to synthetic Planck data.}
For estimating the power spectra from synthetic Planck data, we use the nonparametric regression method described in \citep{AAS2012}.
While our synthetic data is generated under the assumption that the $\Lambda$CDM model as estimated from the WMAP 7-year data is the true model of the Universe,
this formalism for nonparametric regression and inference itself does not make any assumptions about the shape of the true regression function underlying the data; it is asymptotically model-independent.
A nonparametric fit, under this methodology, can be characterized by its effective degrees of freedom (EDoF), which can be thought of as the equivalent of the number of parameters in a parametric regression problem.
Using this methodology, we obtain nonparametric fits to the synthetic $TT$, $EE$ and $TE$ data by appropriately constraining the EDoF of the fits.
(Additional details about our nonparametric fits can be found in Sec.\ \ref{further_details_1} and \ref{further_details_2}.)
Figures \ref{fig_TT_comp}, \ref{fig_EE_comp} and \ref{fig_TE_comp} show nonparametric fits (red curves) to the $TT$, $EE$, and $TE$ data respectively,
which are in good agreement with the underlying $\Lambda$CDM spectra ($C_l^{true}$, black curves) used to generate the synthetic data.
This shows that this nonparametric regression methodology, which does not assume any specific form of the true (but generally unknown) regression function, can recover the underlying true spectrum with high accuracy especially where noise levels are not too high.

\paragraph{How well will the Planck fits be determined by data alone?}
To see how noise in the data affects local uncertainties in a fitted spectrum,
we compute approximate 95\% confidence intervals for each fitted $C_l$ using 5000 randomly sampled spectra from the corresponding confidence set.
The ratio of this confidence interval to the absolute value of the fitted $|C_l|$ (assumed to be nonzero) is a relative measure of how well each fitted $C_l$ is determined \citep{GMN+2004,AAS2012}:
A value $\ll 1$ implies that the fit is well determined by the data, and a value $\gtrsim$ 1 implies that the data contain very little information about height of the power spectrum at that $l$.
 \begin{figure}[t]
  \centerline{
  \includegraphics[height=0.4\textheight]{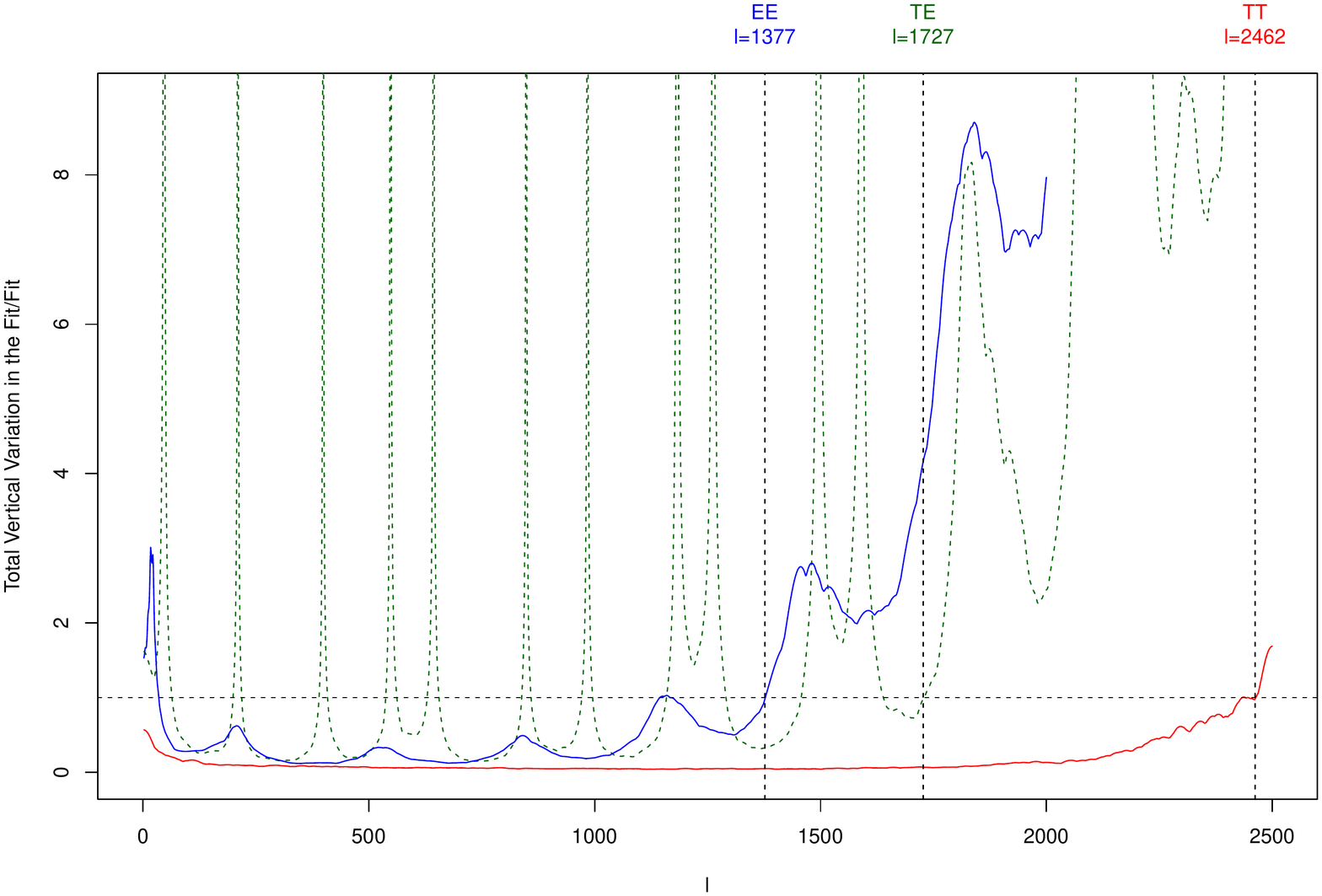}
  }
  \caption{\label{fig_boxcar_planck} The results of a probe of the confidence sets for the $TT$ (red), $EE$ (blue), and $TE$ (green) nonparametric restricted-freedom fits to the synthetic Planck data, to determine how well the fits are expected to be determined by the data alone.
The quantity plotted for each data realization is the total vertical variation at each $l$ within the respective 95\% ($2 \sigma$) confidence set, divided by the absolute value of the fit (assumed nonzero):
Values $\ll 1$ indicate that the fit is tightly determined by the data, whereas values $\gtrsim$ 1 indicate that the data contain very little information about the height of the angular power spectrum at that $l$.
Disregarding the low-$l$ region for the $EE$ fit, and spikes for the $TE$ fit (which arise from nearly zero fitted $C_l$ values), the marked vertical lines indicate the approximate $l$-value at which each curve rises above 1.}
 \end{figure}
In Figure \ref{fig_boxcar_planck}, we plot this relative error (95\%) for all three spectra as a function of the multipole index $l$.
We see that, by this criterion, the Planck power spectra are expected to be well-determined up to $l \approx 2462 (TT), 1377 (EE)$ and $1727 (TE)$.
Since the $TE$ fit oscillates around zero (Figure \ref{fig_TE_comp}), this quantity takes very high values at $l$s where the $TE$ spectrum has a nearly zero value.
This results in multiple spikes in Figure \ref{fig_boxcar_planck} (green dash curve), but this does not imply that the fit is ill-determined at these $l$s.
Ignoring these spikes, we see that the relative error in the $TE$ fit is below unity up to $l \approx 1727$, which indicates the range over which this fit is expected to be well-determined by data alone.

\begin{figure}[h]
  \centerline{\includegraphics[height=0.4\textheight]{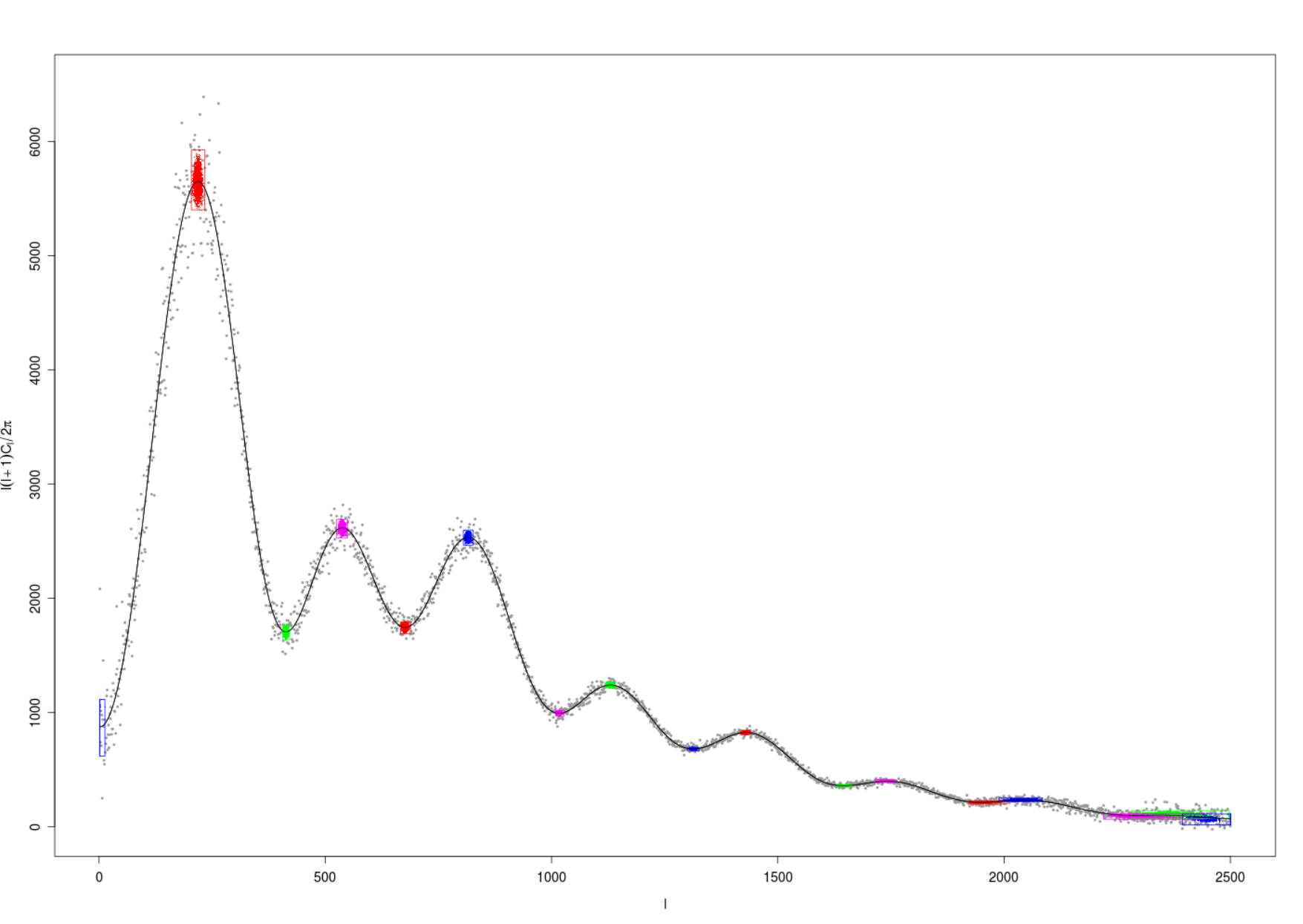}}
  \caption{\label{fig_TT_box} 95\% confidence boxes the locations and heights of peaks and dips in the $TT$ fit. Black curve is the restricted-freedom monotone fit to the synthetic Planck $TT$ data (grey points). The number of acceptable spectral variations sampled from the 95\% confidence set is 5000. These uncertainties are tabulated in Table \ref{table:TT}.}
 \end{figure}

\begin{figure}[h]
  \centerline{\includegraphics[height=0.4\textheight]{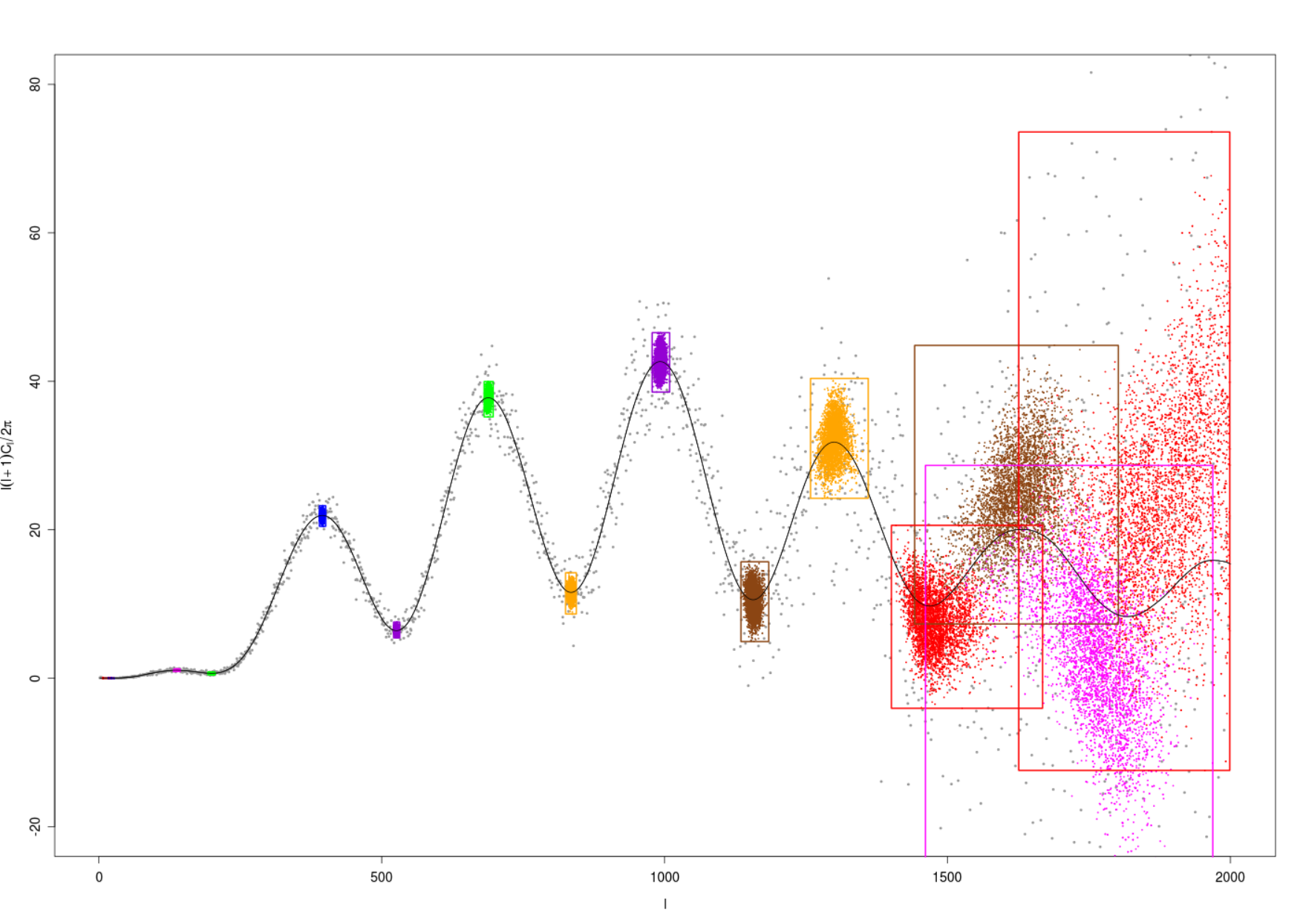}}
  \caption{\label{fig_EE_box} 95\% confidence boxes the locations and heights of peaks and dips in the $EE$ fit. Black curve is the restricted-freedom monotone fit to the synthetic Planck $TT$ data (grey points). The number of acceptable spectral variations sampled from the 95\% confidence set is 5000. These uncertainties are tabulated in Table \ref{table:EE}.}
 \end{figure}

\begin{figure}[h]
  \centerline{\includegraphics[height=0.4\textheight]{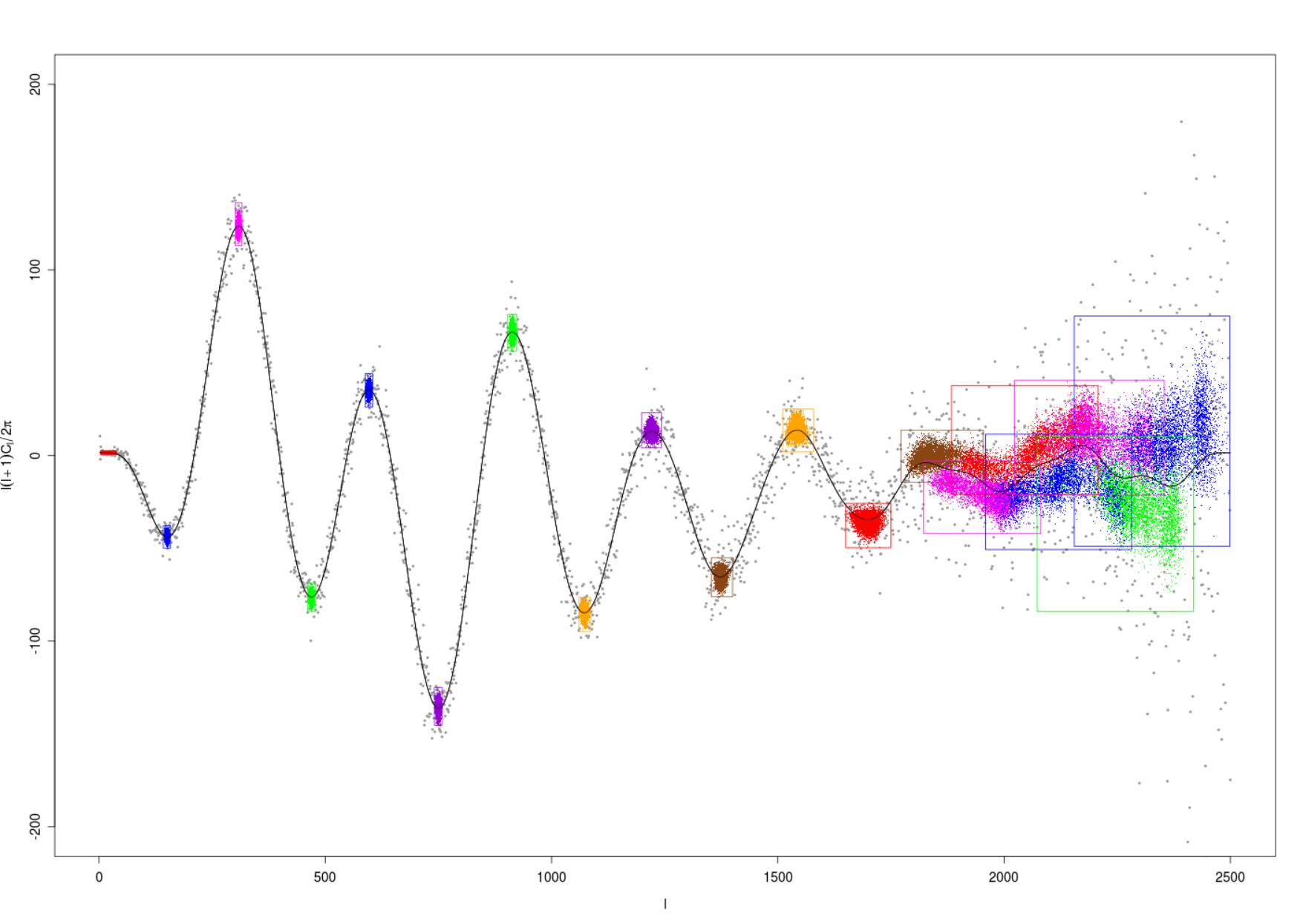}}
  \caption{\label{fig_TE_box} 95\% confidence boxes the locations and heights of peaks and dips in the $TE$ fit. Black curve is the restricted-freedom monotone fit to the synthetic Planck $TT$ data (grey points). The number of acceptable spectral variations sampled from the 95\% confidence set is 5000. These uncertainties are tabulated in Table \ref{table:TE}.}
 \end{figure}

\paragraph{Uncertainties on the locations and heights of peaks and dips.}
Locations and heights of peaks and dips in the CMB power spectra contain information about cosmological models and parameters.
Uncertainties in the location and height of a peak or a dip in a fitted spectrum can thus help assess uncertainties in the values of related parameter.
Following the procedure outlined in \citep{AAS2012}, we sampled the 95\% confidence set of each fitted spectrum uniformly to generate spectral variations while ensuring that at least 5000 of these are acceptable (see Sec.\ \ref{further_details_2} for details).
Figures \ref{fig_TT_box}, \ref{fig_EE_box}, and \ref{fig_TE_box} show the results of this exercise, together with tabulated values in Tables \ref{table:TT}, \ref{table:EE}, and \ref{table:TE} respectively.
The box around a peak or a dip represents the largest horizontal and vertical variations in the scatter; these represent the 95\% confidence intervals on the location and height of a peak or a dip.

For the $TT$ fit, these boxes around peaks and dips are tiny (Figure \ref{fig_TT_box}), which is a reflection of the accuracy of the Planck $TT$ data.
Such precise determination of peaks and dips should lead to more robust estimates of related cosmological parameters than what is currently available.
The theoretical $TT$ power spectrum (Figure \ref{fig_TT_comp}, black curve) shows a small upturn at low $l$.
This upturn is primarily the result of the integrated Sachs-Wolf (ISW) effect.
In Figure \ref{fig_TT_box}, this upturn corresponds to the first dip in the spectral variations sampled from the 95\% confidence set.

Peaks and dips in the $EE$ power spectrum show reasonably low uncertainties up to $l \approx 1200$ (Figure \ref{fig_EE_box}).
Beyond this, the data contain high levels of noise, and therefore all uncertainties become much larger.
This is in agreement with the behavior of the $EE$ curve in Figure \ref{fig_boxcar_planck} (blue curve).
We also expect a small bump in the $EE$ power spectrum which is related to the epoch of reionization.
This bump is indeed seen in both nonparametric $EE$ fits in Figure \ref{fig_EE_comp}.
Figure \ref{fig_EE_box} shows a (tiny) 95\% uncertainty box for this bump.
A precise determination of the height and location of this peak (Table \ref{table:EE}) should reveal useful information about the epoch of reionization.

The uncertainty boxes on peaks and dips in the $TE$ fit (Figure \ref{fig_TE_box}) are reasonably small till $l\approx 1800$, again in agreement with the result depicted in Figure \ref{fig_boxcar_planck} (green dashed curve).
The somewhat peculiar uncertainty boxes on the last three peaks in the $TE$ fit
are due to the fact that there are two spurious peaks at $l = 1920$ and $l = 2070$ in the restricted-freedom fit (see Sec.\ \ref{further_details_1}) which are a result of the high noise levels at high $l$s.
At low multipoles, the $TE$ fit also shows a bump  (Figure \ref{fig_TE_comp}) which is related to the epoch of reionization.
Although a similar bump in the $EE$ fit is known to be more informative \citep{TPC2006}, a determination of this peak in the $TE$ spectrum should also lead to useful information about reionization.

\begin{figure}[th]
  \centerline{
  \includegraphics[width=0.65\textwidth]{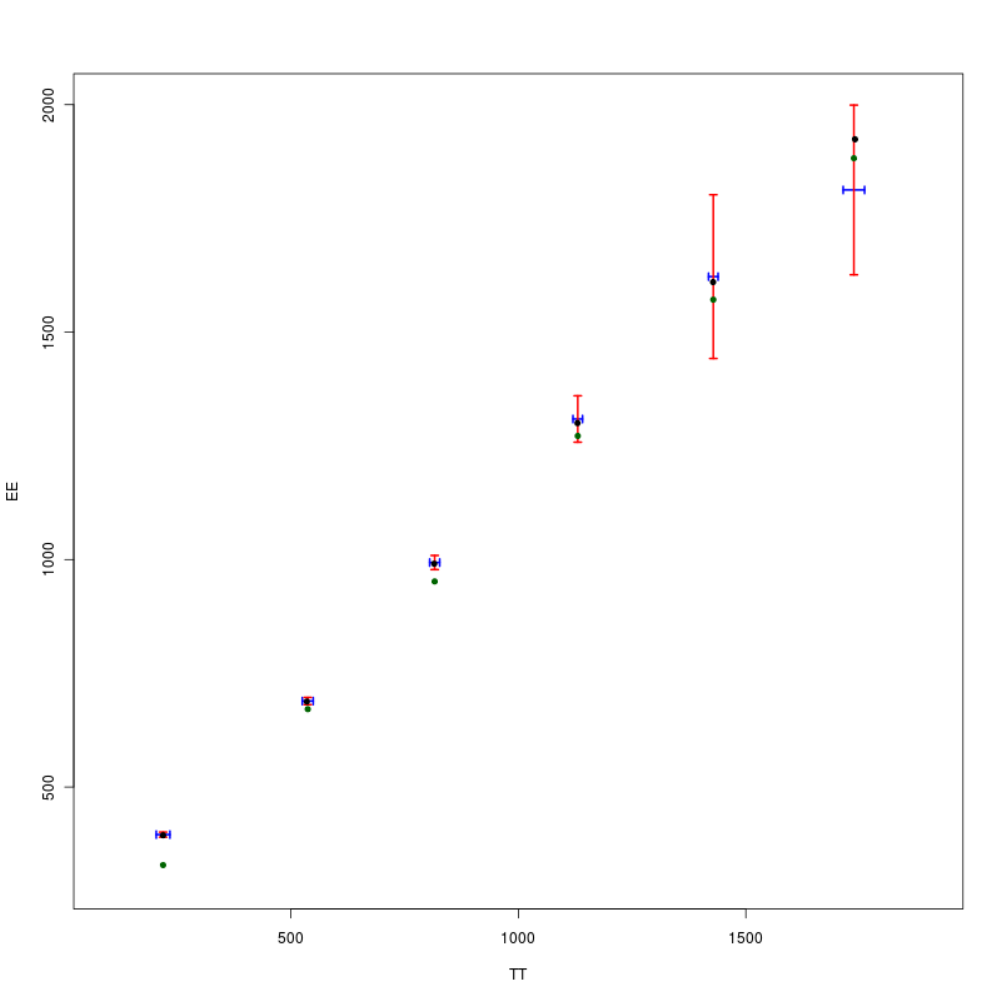}
  }
  \caption{\label{fig_EE_TT_ratio_Planck} The 95\% confidence intervals on peak locations in the $EE$ fit (red) plotted against the corresponding confidence intervals in the $TT$ fit (blue). Also plotted are the peak location pairs corresponding to the best-fit $\Lambda$CDM model (black dots), and points (green) based on the theoretical expectation $(m+0.5)/m$. All the plotted quantities are by and large consistent with each other, indicating that the expected behavior of out-of-phase peak locations is indeed vindicated by the data.}
 \end{figure}

\paragraph{Are the acoustic peaks in the $TT$ and $EE$ spectra out of phase with respected to each other?}
From the fundamental physics of the CMB anisotropies, we expect the acoustic peaks in the $TT$ and $EE$ power spectra to be out phase with respect to each other.
One way of establishing this is by considering the ratio of peak locations in the $EE$ fit to the corresponding ones in the $TT$ fit, which should be
$(m+0.5)/m$ for the $m$th peak \citep{Sievers2007}.
We depict the 95\% peaks locations of TT power spectra versus EE one in
Figure \ref{fig_EE_TT_ratio_Planck} shows the 95\% confidence intervals on peak locations in the $EE$ fit (red) plotted against the corresponding uncertainties in the $TT$ fit (blue).
Also plotted are the peak location pairs corresponding to the best-fit $\Lambda$CDM model (black dots),
and points (green) based on the theoretical expectation $(m+0.5)/m$.
All the plotted quantities are by and large consistent with each other,
indicating that the expected behavior of out-of-phase peak locations is indeed vindicated by the data.



\begin{figure}[h]
  \centerline{\includegraphics[width=\textwidth]{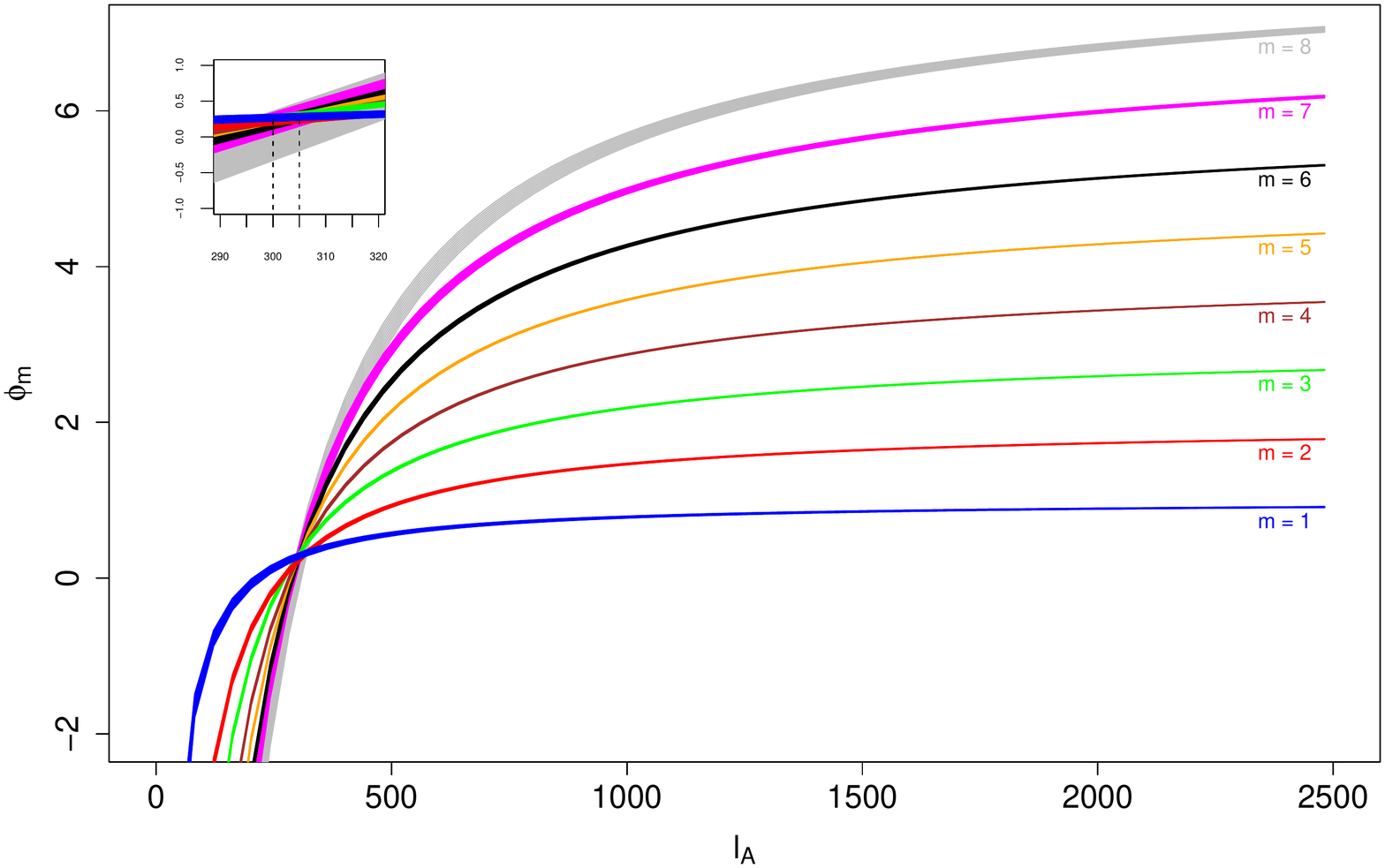}}
  \caption{\label{fig_phi_vs_lA_8_peak} Confidence ``bands'' for the acoustic scale $l_A$ and the shift $\phi_m$ for the $m$th peak, as derived from the 95\% confidence intervals on the first eight peak locations of estimated TT power spectrum.} 
 \end{figure}

\paragraph{An estimate of the acoustic scale parameter $l_A$.}
Table \ref{table:TT} lists the 95\% confidence intervals on peak and dip locations and heights for the $TT$ power spectrum fit.
As a way of illustrating the role of these uncertainties in the estimation of cosmological parameters, we consider the following relationship \citep{HFZ+2001,DL2002} between the location $l_m$ of the $m$th peak, the acoustic scale $l_A$, and the shift parameter $\phi_m$ for TT power spectrum: $l_m = l_A ( m - \phi_m )$.
If we substitute the end-points of the 95\% confidence interval for the $m$th peak location, then
this relationship results into hyperbolic confidence bands in the $l_A-\phi_m$ plane (Figure \ref{fig_phi_vs_lA_8_peak}).
The intersection of these bands (for the first 8 peaks in the $TT$ fit) determine an estimated confidence interval for the acoustic scale $300\leq l_A \leq 305$ which is in agreement with the reported value $l_A = 300$ by \citep{PNB+2003}.
In comparison with our previous estimate based on the WMAP 7-year data \citep{AAS2012}, the current estimate has improved remarkably.
Furthermore, any additional information about the phase shifts $\phi_m$ can lead to an even more refined estimate for acoustic scale.

\section{Conclusion}
\label{conclusion}

In this paper, we have addressed the question of what could be the expected outcome of a nonparametric analysis of the $TT$, $EE$ and $TE$ power spectrum data sets from the Planck space mission when they get released.
For this purpose, we have used synthetic/simulated Planck-like data sets based on the best-fit $\Lambda$CDM model,
and have analysed them using a nonparametric regression and inference methodology.
Our results show that the $TT$ power spectrum can be estimated with such a high accuracy that all peaks are resolved up to $l\leq2500$.
We expect that the $EE$ and $TE$ power spectra will be reasonably well-determined and can help, e.g., better understand reionization history, address the issue of adiabatic versus isocurvature perturbations, and estimate of some of the cosmological parameters precisely.
As a result, we expect to have a better understanding of the Universe via the Planck data even from an agnostic, nonparametric approach.


\begin{table}[h]
\centering
\scalebox{0.9}{
\begin{tabular}{l|l|l|l}

Peak Location    &      Peak Height            &        Dip Location                 & Dip Height    \\
\hline
\hline

$l_1: (204, 234)$   &  $h_1: (5401.543, 5925.611)$    &   $l_{1+{1 \over 2}}: (406, 421)$    &   $h_{1+{1 \over 2}}: (1638.521, 1769.197)$ \\
$l_2: (525, 549)$   &  $h_2: (2527.018, 2691.862)$    &   $l_{2+{1 \over 2}}: (666, 688)$    &   $h_{1+{1 \over 2}}: (1689.753, 1793.608)$ \\
$l_3: (805, 827)$   &  $h_3: (2465.483, 2596.654)$    &   $l_{3+{1 \over 2}}: (1009, 1028)$  &   $h_{1+{1 \over 2}}: (969.1924, 1019.3422)$ \\
$l_4: (1120, 1141)$ &  $h_4: (1215.164, 1267.128)$    &   $l_{4+{1 \over 2}}: (1301, 1325)$  &   $h_{1+{1 \over 2}}: (663.6103, 695.9041)$ \\
$l_5: (1418, 1439)$ &  $h_5: (806.5287, 845.0737)$    &   $l_{5+{1 \over 2}}: (1630, 1666)$  &   $h_{1+{1 \over 2}}: (348.6765, 370.1998)$ \\
$l_6: (1714, 1761)$ &  $h_6: (387.0516, 412.0869)$    &   $l_{6+{1 \over 2}}: (1922, 2007)$  &   $h_{1+{1 \over 2}}: (199.8328, 225.3753)$ \\
$l_7: (1989, 2085)$ &  $h_7: (219.8173, 251.4042)$    &   $l_{7+{1 \over 2}}: (2220, 2442)$  &   $h_{1+{1 \over 2}}: (64.88648, 112.57429)$ \\
$l_8: (2282, 2498)$ &  $h_8:  (71.91322, 141.91509)$  &   $l_{8+{1 \over 2}}: (2394, 2500)$  &   $h_{1+{1 \over 2}}: (15.27434, 111.02498)$ \\

\hline
\end{tabular}
}
\caption{95\% Confidence Interval on Several Features of TT Angular Power Spectrum}
\label{table:TT}
\end{table}
\begin{table}[h]
\centering
\scalebox{0.9}{
\begin{tabular}{l|l|l|l}

Peak Location    &      Peak Height            &        Dip Location                 & Dip Height    \\
\hline
\hline
$l_1: (7, 26)$      &  $h_1: (-0.0025,0.0235)$      &   $l_{1+{1 \over 2}}: ( 17, 27)$     &   $h_{1+{1 \over 2}}: (-0.0029,  0.0181)$ \\
$l_2: (133, 144)$   &  $h_2: ( 0.9234, 1.2542)$     &   $l_{2+{1 \over 2}}: ( 193, 205)$   &   $h_{2+{1 \over 2}}: (0.4620, 0.8600)$\\
$l_3: (390, 401)$   &  $h_3: ( 20.4654, 23.2528)$   &   $l_{3+{1 \over 2}}: ( 521, 532)$   &   $h_{3+{1 \over 2}}: (5.4424, 7.5745)$\\
$l_4: (681, 697)$   &  $h_4: ( 35.2102, 39.9650)$   &   $l_{4+{1 \over 2}}: ( 825, 844)$   &   $h_{4+{1 \over 2}}: (8.6629, 14.2180)$\\
$l_5: (978, 1009)$  &  $h_5: ( 38.5544, 46.5531)$   &   $l_{5+{1 \over 2}}: ( 1135, 1184)$ &   $h_{5+{1 \over 2}}: (4.9459, 15.6909)$\\
$l_6: (1258, 1360)$ &  $h_6: ( 24.2445, 40.3917)$   &   $l_{6+{1 \over 2}}: ( 1401, 1668)$ &   $h_{6+{1 \over 2}}: (-4.0435, 20.6010)$\\
$l_7: (1442, 1802)$ &  $h_7: (  7.3139, 44.8427)$   &   $l_{7+{1 \over 2}}: ( 1461, 1969)$ &   $h_{7+{1 \over 2}}: (-33.4637,  28.6720)$\\
$l_8: (1626, 1999)$ &  $h_8: ( -12.4126,  73.5849)$ &   $l_{8+{1 \over 2}}: ( 1755, 2000)$ &   $h_{8+{1 \over 2}}: (-46.4440,  65.6556)$\\

\hline

\end{tabular}
}
\caption{95\% Confidence Interval on Several Features of EE Angular Power Spectrum}
\label{table:EE}
\end{table}
\begin{table}[h]
\centering
\scalebox{0.9}{
\begin{tabular}{l|l|l|l}

Peak Location    &      Peak Height            &        Dip Location                 & Dip Height    \\
\hline
\hline
$l_1: (3, 38)$       &  $h_1: (0.6092, 2.5822)$      &   $l_{1+{1 \over 2}}: ( 142, 158)$   &   $h_{1+{1 \over 2}}: (-49.8581, -37.7906)$ \\
$l_2: (301, 316)$    &  $h_2: (113.1289, 136.2571)$  &   $l_{2+{1 \over 2}}: (461, 477)$    &   $h_{2+{1 \over 2}}: (-83.9852, -68.6596)$\\
$l_3: (588, 605)$    &  $h_3: (26.2370, 44.0873)$    &   $l_{3+{1 \over 2}}: (741, 758)$    &   $h_{3+{1 \over 2}}: (-145.3450, -124.9381)$\\
$l_4: (903, 923)$    &  $h_4: (56.4278, 76.1789)$    &   $l_{4+{1 \over 2}}: (1061, 1086)$  &   $h_{4+{1 \over 2}}: (-94.7879, -76.2354)$\\
$l_5: (1199, 1243)$  &  $h_5: (4.2335, 23.1510)$     &   $l_{5+{1 \over 2}}: (1353, 1400)$  &   $h_{5+{1 \over 2}}: (-76.0768, -55.1005)$\\
$l_6: (1511, 1579)$  &  $h_6: (1.9257, 25.1787)$     &   $l_{6+{1 \over 2}}: (1650, 1750)$  &   $h_{6+{1 \over 2}}: (-49.6416, -25.6796)$\\
$l_7: (1772, 1954)$  &  $h_7: (-14.4294,  13.7468)$  &   $l_{7+{1 \over 2}}: (1822, 2081)$  &   $h_{7+{1 \over 2}}: (-42.0202,  -2.8288)$\\
$l_8: (1884, 2208)$  &  $h_8: (-20.8813,  37.7615)$  &   $l_{8+{1 \over 2}}: (1959, 2282)$  &   $h_{8+{1 \over 2}}: (-50.6428,  11.4636)$\\
$l_9: (2023, 2354)$  &  $h_9: (-21.1805,  40.5083)$  &   $l_{9+{1 \over 2}}: (2073, 2419)$  &   $h_{9+{1 \over 2}}: (-83.8615,  10.2379)$\\
$l_10: (2155, 2499)$ &  $h_10: (-48.9460,  75.1592)$ &   $l_{10+{1 \over 2}}: (2257, 2500)$ &   $h_{10+{1 \over 2}}: (-116.2237,   60.3390)$\\
\hline

\end{tabular}
}
\caption{95\% Confidence Interval on Several Features of TE Angular Power Spectrum}
\label{table:TE}
\end{table}

\appendix

\section{Full-freedom and restricted-freedom nonparametric fits}
\label{further_details_1}

The full-freedom fits are obtained by minimizing the risk function subject to monotonicity constraints on the shrinkage parameters \citep{AAS2012}.
Such full-freedom fits can be quite oscillatory especially where the noise levels in the data are high.
Despite this fit being a reasonable fit (in the sense that it captures the essential trend in the data well), all cosmological models expect far smoother angular power spectra.
To account for this, we minimize the risk again, restricting the EDoF of the fit to a value less than that for the full-freedom fit.
We continue reducing the EDoF in this way until we obtain an acceptably smooth fit.
We call this the \emph{restricted-freedom fit}.

The full-freedom fit for the $TT$ power spectrum is almost smooth, but with tiny wiggles (Figure \ref{fig_TT_comp}, blue curve), and corresponds to EDoF$\approx 72$.
We achieve a numerically smooth fit for the $TT$ power spectrum at EDoF$\approx 27$ (Figure \ref{fig_TT_comp}, red curve).
We see that this restricted-freedom fit follows the full-freedom fit almost exactly, but without the additional wiggles which resulted from noise in the data.
For the $EE$ power spectrum, the full-freedom fit has EDoF$\approx 189$ (Figure \ref{fig_EE_comp}, blue curve).
This fit is quite wiggly at high multipoles, as can be expected from high noise levels in data there.
The Planck proposal \citep{TPC2006} also expects the $EE$ power spectrum to have high noise at high multipoles, and hence unable to resolve peaks beyond $l \lesssim 1000$.
Therefore, we have excluded the $EE$ fit beyond $l=2000$ (Figure \ref{fig_EE_comp}) from further analysis.
The restricted-freedom smoother version of the $EE$ fit corresponds to EDoF$=23.79$ (Figure \ref{fig_EE_comp}, red curve).
We see again that the smooth fit follows the full-freedom fit, but averages over the wiggles.
For the simulated $TE$ data, we obtain the full-freedom fit at EDoF$\approx 95$ (Figure \ref{fig_TE_comp}, blue curve).
The peaks are resolved well up to $l \approx 2000$.
At higher multipoles, we see a wiggly fit with false peaks due to high levels of noise.
A smoother restricted-freedom fit is obtained at EDoF$=40$ (Figure \ref{fig_TE_comp}, red curve).
Despite this additional smoothing, because of the high noise at high multipoles,
we see two false bumps at $l=1920,2070$, and a false peak at $l=2304$.

\section{Probing the confidence sets}
\label{further_details_2}

A high-dimensional confidence set (for a prespecified level of confidence; typically, 95\%) around a fit is the prime inferential object for this nonparametric methodology \citep{AAS2012}.
By probing the confidence set, we can determine uncertainties on specific feature of a fit, validate different cosmological models against the data, etc.
For finding uncertainties on the location and height of a peak or a dips (Figures \ref{fig_TT_box}, \ref{fig_EE_box}, and \ref{fig_TE_box})
we use the same methodology as was used in \citet{AAS2012}:
We uniformly sample the confidence set of each smooth fit, and record the peaks and dips of acceptable spectra thus sampled.
The most extreme variations in the height or location define a confidence interval on the respective estimates (represented in the figures as boxes).
Compared to \citet{AAS2012}, we change a few criteria for accepting such sampled spectra to account for the greater angular resolution of the Planck data:
For the $TT$ and $TE$ fits, we select spectra with exactly 8 and 10 peaks respectively.
In the case of the $TE$ fit, this criterion includes a peak at low multipoles ($l\leq 50$) and a false peak at $l=2304$.
For the $EE$ fit, we sample spectra with 8 peaks for $l \le 2000$, together with the additional condition that there must be a peak at low multipoles ($l\leq 50$).
The last peak in the $EE$ restricted-freedom fit consists of two tiny but close bumps due to noise in the data.
Such fine structure is not expected here on theoretical grounds.
Therefore, in case we find a sampled spectrum with two tiny peaks around the location of the eighth peak in the fit, but with separation less than 10 multipoles, we consider them as a single peak and record their average location and height.

\newpage

\bibliographystyle{apj}
\bibliography{cmb,cosmomc}

\end{document}